# Chiral 480nm absorption in the hemoglycin space polymer.

Julie E M McGeoch[1] and Malcolm W McGeoch[2]

[1]High Energy Physics DIV, Smithsonian Astrophysical Observatory, Center for Astrophysics, Harvard & Smithsonian, 60 Garden St, MS 70, Cambridge MA 02138, USA.
[2]PLEX Corporation, 275 Martine St, Suite 100, Fall River MA 02723, USA.

**Abstract**
A 1494 Dalton hemoglycin space polymer of Glycine$_{18}$ Hydroxy-glycine$_4$ Fe$_2$O$_4$ termed the "core unit" is part of a polymer of Glycine, Si, Fe and O that forms tubes, vesicles and a lattice structure isolated from CV3 meteorites and characterized by mass spectrometry, FIB/SIMS and X-ray analysis. In Hartree-Fock calculations the polymer has an absorption of blue light at 480nm that is dependent on rectus "R" (= dextro D) chirality in a hydroxy-glycine residue whose C-terminus is bonded to an iron atom. The absorption originates in the Fe II state as a consequence of chiral symmetry breaking. The infrared spectrum is presented. We discuss how the core unit could have been selected 4.5 billion years ago in our protoplanetary disc by blue light from the early sun.

**Introduction**
The structural meteoritic polymers discovered and characterized in our prior work are here modeled at a high level of quantum chemistry to predict their optical absorptions for astronomers. Being widespread in carbonaceous chondrites of the early CV3 type and having $^2$H and $^{15}$N isotope levels characteristic of comets, these molecules could have played a role in the accretion of our solar system and therefore might be observable in other protosolar or circumstellar discs. The molecular structure of these space polymers, that we term in general hemoglycin, contains iron atoms that close the ends of anti-parallel polyglycine chains in a previously unknown iron configuration.

The core unit (Glycine$_{18}$ Hydroxy-glycine$_4$ Fe$_2$O$_4$) and its associated polymers have been determined by extracting the molecule from micron particles of CV3 meteorite samples and then analyzing the extracts by mass spectrometry [1,2,3,4] and X-ray diffraction [5] with supporting measurements of $^{15}$N [6] via FIB/SIMS. The MALDI mass spectrometry technique basically provides data on an intact molecule and/or its fragments with data displayed as peak amplitude versus mass to charge ratio (*m/z*). Thousands of MALDI peaks were analyzed [1,2,3,4] allowing models of the polymer structure to be built by software aided by X -ray diffraction data of crystals of the polymer [5]. The polymer forms rolled up tubes, can cover a surface forming spherical vesicles, and self-organizes into a three-dimensional low-density lattice [5] that is depicted in Figure 1. The source meteorite samples were Allende [1,2,3], Acfer-086 [2,3,4,5] and Kaba [4,5] while Efremovka and Sutter's Mill extracts both showed macroscopic evidence of the polymer structures.

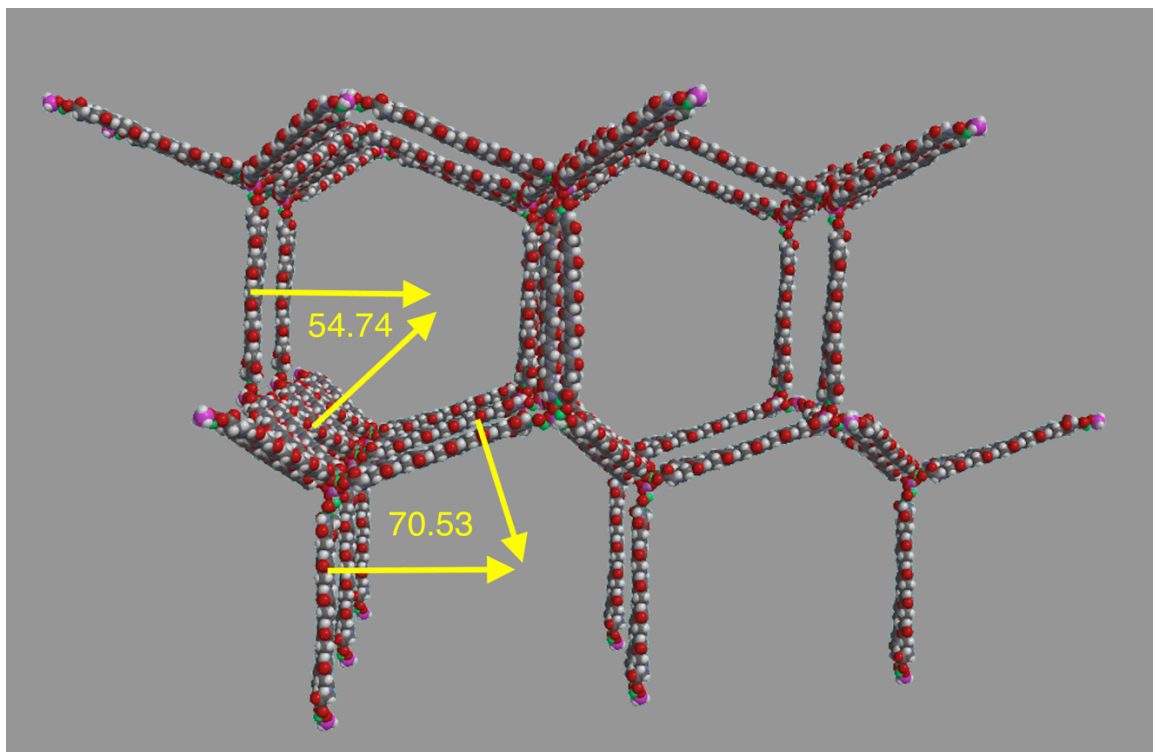

**FIGURE 1: Lattice form of the polymer [5] built of modified core units, here at 1638Da, linked by silicon atoms. The tetrahedral angles have been confirmed in X-ray scattering. Space filling model. Atoms: hydrogen white, carbon black, nitrogen blue, oxygen red, iron green, silicon pink.**

UV/visible and IR calculations were run on the core protein and on pared down versions of that structure using equilibrium geometry, Hartree-Fock, 3-21G. Absorption wavelengths and strengths were calculated to 20 excited states, which covered the range down to the mid-ultraviolet. Transition strengths were typically plotted on a $\log_{10}$ vertical scale unless mentioned otherwise, and an artificial width of 40nm was applied to the spectral peaks to simulate the typical molecular broadening via vibrations. The calculations were all in gas phase, which we believe is more appropriate for molecules in molecular clouds or discs. The pared down structures had less glycine units per anti-parallel beta sheet chain and beyond the 2Fe case included a core with only a single Fe, loops of glycine with no Fe, the core minus OH groups, different placements of the OH groups and changes to the chirality of the OH groups.

**Results**
The "core unit", comprises two poly-glycine chains of length 11 residues in an anti-parallel beta sheet with iron atoms joining the chains at each end [4]. Four of these glycine residues have OH groups on the alpha carbon units, in 'S' or 'R' chirality (Laevo- or Dextro- rotatory, respectively). An absorption at 480nm is induced only when there is an 'R' chirality hydroxy-glycine connected via its peptide C-terminal to a terminal iron atom. The absorption is absent in all other cases involving (plain) glycine or 'S' hydroxy-glycine at either the C-terminal or the N-terminal adjacent to an Fe atom, also it is absent when 'R' hydroxy-glycine has its N-terminal next to iron. The origin of this induced optical transition lies in the symmetry-breaking that occurs when an otherwise symmetrical iron atom is attached via covalent bonds to an asymmetric polypeptide structure with directionality, i.e. the iron orbitals are constrained from rotating relative to the polypeptide structure, which itself has chirality. We find that the absorption relates to the Fe II ($Fe^+$) state of iron in specific molecular surroundings.

Mass spectrometry had revealed that an even number of glycine residues was contained in the dominant m/z peaks, suggesting that there could be a pairing of equal length single strands of polyglycine. Energy minimization pointed to an anti-parallel pairing owing to the strength of the {C=O ::: H-N} hydrogen bonds between strands in that orientation. This structure, referred to as polyglycine I, is fully described by Lotz [7] and Moore and Krimm [8] in their analysis of its infrared (IR) spectrum vibrations. Each "core unit" chain has 11 glycine residues with four of these (two at each end) modified into hydroxy-glycine ($Gly_{OH}$). The chains are linked at the ends via Fe atoms. Two possible versions of the core unit, differing by chirality, are shown in Figure 2 in equilibrium geometry. Our earlier mass spectrometry work indicated via study of fragments that the hydroxy-glycine residues were very probably adjacent to terminal Fe atoms, but mass spectrometry alone cannot determine residue chirality.

One example of chiral induction of a 480nm absorption is shown in Figure 2 in which the vertical scale represents calculated transition strengths on a $log_{10}$ scale and the chirality of $Gly_{OH}$ is indicated by 'S' or 'R'. The range of calculations covered 5, 7, 9, 11 and 13 residue chains, with two, one or zero Fe atoms. When Fe was absent the polymer was given an additional peptide bond at that location, to make a continuous loop. The cases without any Fe had only a main deep ultraviolet absorption in the region of 150-200nm (S1 Fig.1) and did not exhibit any of the visible or near infrared absorptions discussed here. The latter were associated only with the presence of iron, and furthermore with single iron atoms that could be identified by the Mulliken charge as being at one or the other end of the molecule. To complete calculations on the nine cases that covered zero, 'S'-chiral and 'R'-chiral $Gly_{OH}$ located on either the C-terminus or the N-terminus adjacent to Fe, we reduced the chain length to 7 residues, having first determined that the results did not depend upon chain length from 7 thru 13, but only upon the local arrangement around one iron atom. The calculated absorption wavelengths for these 9 cases are listed in Table 1.

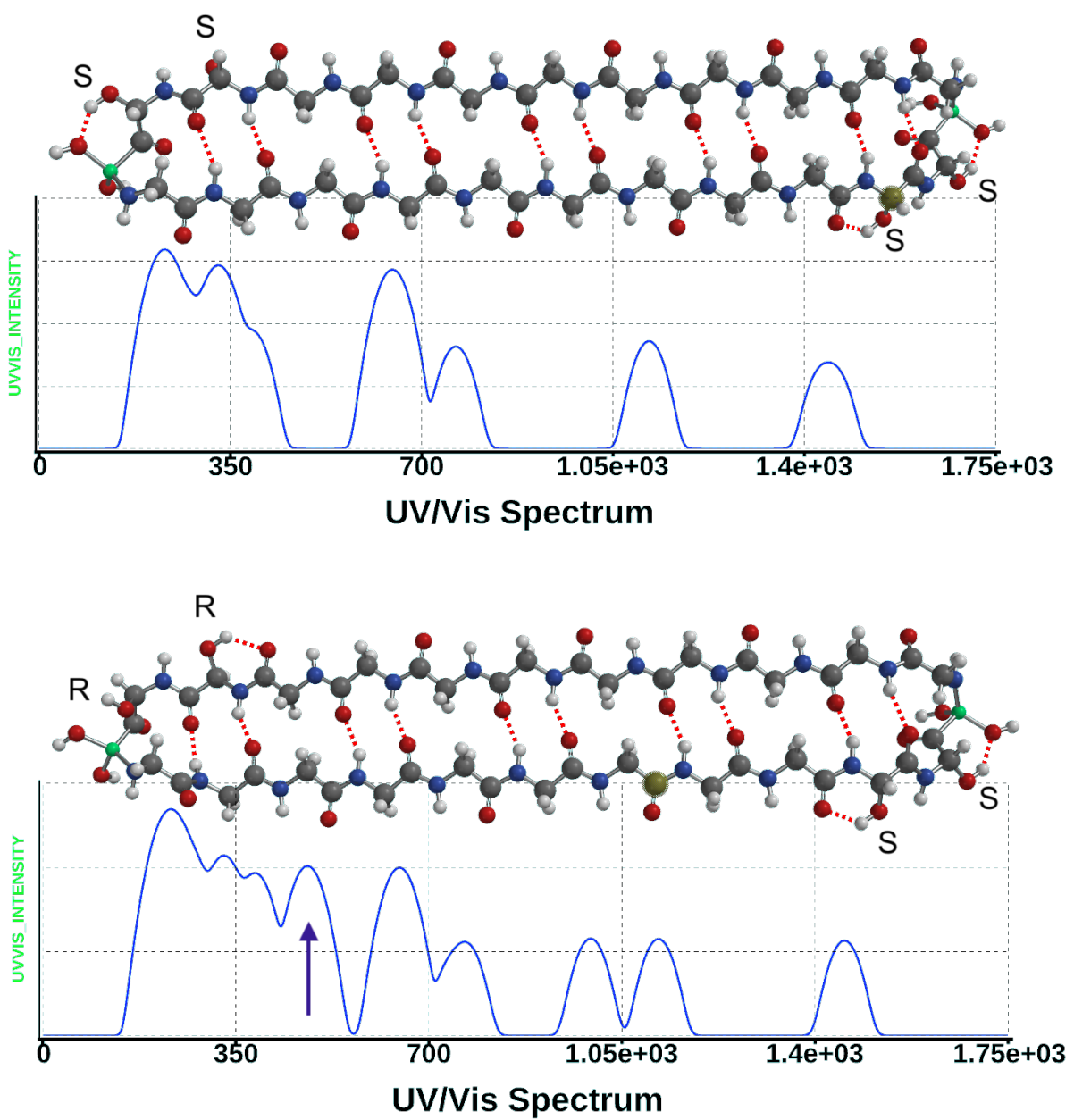

**FIGURE 2a** (top) UV/visible/IR spectra from the core polymer where the 4 hydroxy-glycine units all have 'S' (sinister/laevo S/L) chirality. There is no absorption at 480nm. Graph axes: Vertical; calculated transition strengths on a $\log_{10}$ scale: Horizontal; wavelength (nm). The molecular model format is ball and spoke. Atom labels: hydrogen white, carbon black, nitrogen blue, oxygen red, iron green. **2b** (bottom) spectrum from the core polymer with an 'R' chirality hydroxy-glycine adjacent to an Fe atom at the peptide C-terminal.

Without Gly$_{OH}$, in column 1 of Table 1, the calculated absorptions rise in pairs, which correlate with first one, then the other iron atom. Generally, the lowest energy absorption varies toward longer wavelength with any of the Gly$_{OH}$ residues present (first row of Table 1). The second row relates to the distal iron atom, and hence is immune to environmental changes around the first Fe.

**Table 1.** "7-residue chain" with zero, one or two Gly$_{OH}$ residues at one end only, coded by its absence (0) or location on {N-terminal, C-terminal} and chirality (S or R) via {(0,S,R),(0,S,R)}. For example, {0,R} refers to a C-terminal R Gly$_{OH}$ only, with unmodified glycine at the N-terminal. Wavelengths (nm) and transition strengths are from the RPA calculations.

| State indice | Case 1 {0,0} | 2 {0,S} | 3 {0,R} | 4 {S,0} | 5 {S,S} | 6 {S,R} | 7 {R,0} | 8 {R,S} | 9 {R,R} |
|---|---|---|---|---|---|---|---|---|---|
| 1 | 1221 | 1408 | 1598 | 1252 | 1457 | 1510 | 1406 | 1827 | 1753 |
| 2 | 1221 | 1220 | 1221 | 1221 | 1221 | 1221 | 1216 | 1220 | 1222 |
| 3 | 1045 | 1120 | 1044 | 1045 | 1044 | 1044 | 1041 | 1080 | 1058 |
| 4 | 1044 | 1045 | 1028 | 959 | 1041 | 998 | 1008 | 1044 | 1045 |
| | | | | | | | | | |
| 5 | 757 | 764 | 757 | 816 | 827 | 757 | 762 | 766 | 757 |
| 6 | 757 | 757 | 659 | 757 | 757 | 696 | 753 | 757 | 726 |
| 7 | 596 | 639 | 596 | 596 | 597 | 596 | 596 | 596 | 596 |
| 8 | 596 | 596 | **482** | 556 | 595 | **475** | 530 | 554 | **475** |
| | | | | | | | | | |
| 9 | 361 | 388 | 398 | 365 | 394 | 396 | 370 | 402 | 409 |
| 10 | 361 | 361 | 380 | 361 | 361 | 380 | 362 | 361 | 397 |
| 11 | 347 | 347 | 361 | 347 | 347 | 361 | 361 | 346 | 361 |
| 12 | 346 | 335 | 347 | 346 | 332 | 347 | 347 | 346 | 347 |
| | | | | | | | | | |
| 13 | 331 | 332 | 332 | 332 | 330 | 331 | 333 | 331 | 332 |
| 14 | 331 | 326 | 330 | 332 | 322 | 330 | 331 | 316 | 327 |
| 15 | 291 | 319 | 291 | 293 | 319 | 291 | 292 | 311 | 291 |
| 16 | 291 | 291 | 274 | 291 | 291 | 263 | 282 | 291 | 276 |
| | | | | | | | | | |
| 17 | 254 | 269 | 254 | 254 | 264 | 254 | 254 | 254 | 254 |
| 18 | 254 | 254 | 240 | 250 | 254 | 236 | 243 | 253 | 243 |
| 19 | 226 | 230 | 227 | 226 | 226 | 226 | 226 | 227 | 233 |
| 20 | 226 | 226 | 226 | 218 | 224 | 223 | 223 | 226 | 226 |

**Interpretation of the calculated spectra relative to Fe structure**

In cases 3, 6 and 9 of Table 1, which have an R-Gly$_{OH}$ C-terminal bonded to Fe, we noted that there was a good correlation between the calculated absorption wavelengths (in RPA) and a set of transitions in the Fe II (Fe$^+$) spectrum. The wavelengths of corresponding states in columns 3, 6 and 9 were averaged and plotted in Figure 3 against the $Fe^+$ (Fe II) transitions [9,10,11] listed in Table 2. An exact linear correlation was seen, within which the open circles represented transitions of Fe II that were not allowed due to parity, while the solid circles showed allowed transitions. In Fe II there are two close-lying ground states, the Fe II $a^4F_{9/2}$ quasi-ground state lying only 1,872cm$^{-1}$ above the true ground state $a^6D_{9/2}$. Many of the calculated molecular transition wavelengths of Table 1 could be traced to an atomic "root" in either the $a^4F_{9/2}$ or the $a^6D_{9/2}$ state of Fe II as listed in Table 2. In confirmation of the Fe charge state, all the ground states listed in Table 2 had a Mulliken charge of 1.06 units, effectively acting as $Fe^+$ (Fe II) ions. The calculated strength of the 477.5nm absorption was 0.0116, averaged over cases 3, 6 and 9, which greatly exceeded the strengths of the other parity-forbidden transitions of Table 2, indicating a moderate strength for the induced absorption at 477.5nm. Following absorption, re-radiation from this transition would be expected to have a decay life of about 300ns in the absence of further energy transfer into vibrations of the molecule.

An offset of 2,050±460 cm$^{-1}$ in energy was seen between CIS and RPA levels of calculation, with the CIS energies being higher. We identified this constant energy difference with the 1,872 cm$^{-1}$ difference between the two Fe II ground states, indicating a slight difference of outcome between the two methods. As we were able to correlate the calculated RPA results quite accurately with excitations from the Fe II $a^4F_{9/2}$ quasi-ground state (Table 2) we believe that the RPA results quoted throughout this paper are probably reliable to within a few tens of nm.

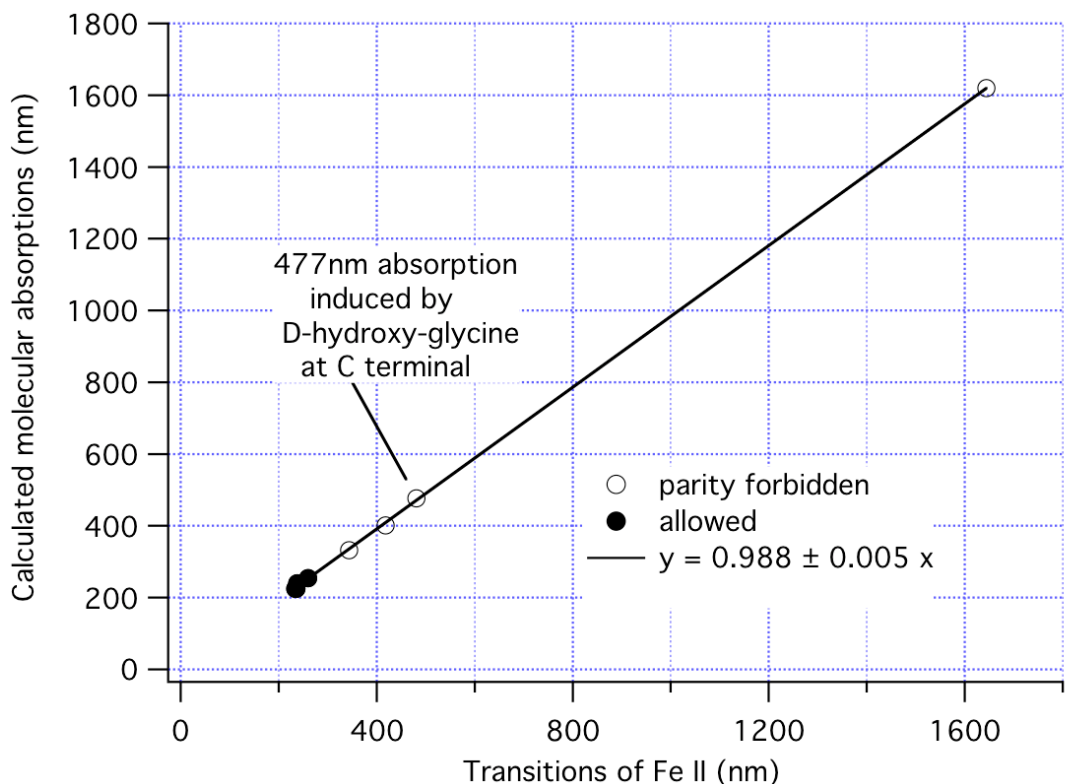

**Figure 3. Correlation between transitions in Fe II ($Fe^+$) and RPA calculated transitions for molecular cases with R(D) chirality hydroxy-glycine in C-terminal contact with Fe.**

**Table 2. List of Fe II transitions either forbidden due to parity (*parity X*) or fully allowed, compared to RPA results for averaged cases 3, 6 and 9 of Table 1 involving R(D) hydroxy-glycine residues with C-termini on Fe.**

| Term g.s. → upper | Energy (cm$^{-1}$) | Wavelength (nm) | Strength | RPA result (nm) | RPA strength |
|---|---|---|---|---|---|
| $a\,^6D_{9/2} \to z\,^6D^O_{9/2}$ | 38,458 | 260 | *allowed* | 254 | 0.0012 |
| $a\,^6D_{9/2} \to z\,^6F^O_{11/2}$ | 41,968 | 238 | *allowed* | 240 | 0.141 |
| $a\,^6D_{9/2} \to z\,^6P^O_{7/2}$ | 42,658 | 234 | *allowed* | 225 | 0.041 |
| | | | | | |
| $a\,^4F_{9/2} \to\ ^4D_{7/2}$ | 6,083 | 1644 | *parity X* | 1620 | 0.00001 |
| $a\,^4F_{9/2} \to b\,^4F_{9/2}$ | 20,765 | 481 | *parity X* | 477.5 | 0.0116 |
| $a\,^4F_{9/2} \to a\,^4G_{11/2}$ | 23,933 | 418 | *parity X* | 401 | 0.00040 |
| $a\,^4F_{9/2} \to b\,^4D_{7/2}$ | 29,053 | 344 | *parity X* | 332 | 0.0071 |
| $a\,^4F_{9/2} \to z\,^4F^O_{9/2}$ | 42,360 | 236 | *allowed* | 228 | 0.062 |

**Infrared spectra**

We calculated the infrared spectra as a guide to astronomical observations of circumstellar discs that could contain space polymers. The infrared (IR) spectra have no major vibration specific to the R/D chirality or positioning of the hydroxy-glycine units. There is a dominant IR vibration in the region of 1651–1658 cm$^{-1}$ (6.0 μm, this calculation, Figure 4) that consists mainly of the C=O stretch mode within two anti-parallel (beta sheet) chains of glycine, named the "amide I" band [8]. However, at 3896 cm$^{-1}$ (2.5μm) there is a vigorous vibration in the Gly$_{OH}$ hydroxyl group (S1 movie Fig.2). This strong movement of the rectus O···H group is a novel finding and may be related to this conformer absorbing UV light at 480nm. The accuracy of these calculations is expected to be better than 5%.

A beta sheet protein backbone gives in general an IR absorption in the region of 6μm and we show this in S1, Figure 3 for the "core unit" compared to the first 10 amino acids in the most conserved of all proteins, subunit c of the ATP synthase: DIDTAAKFIG. Subunit C can have a variety of conformations being the rotor of the mitochondrial ATP synthase complex [12], and a calcium regulated ion channel [13].

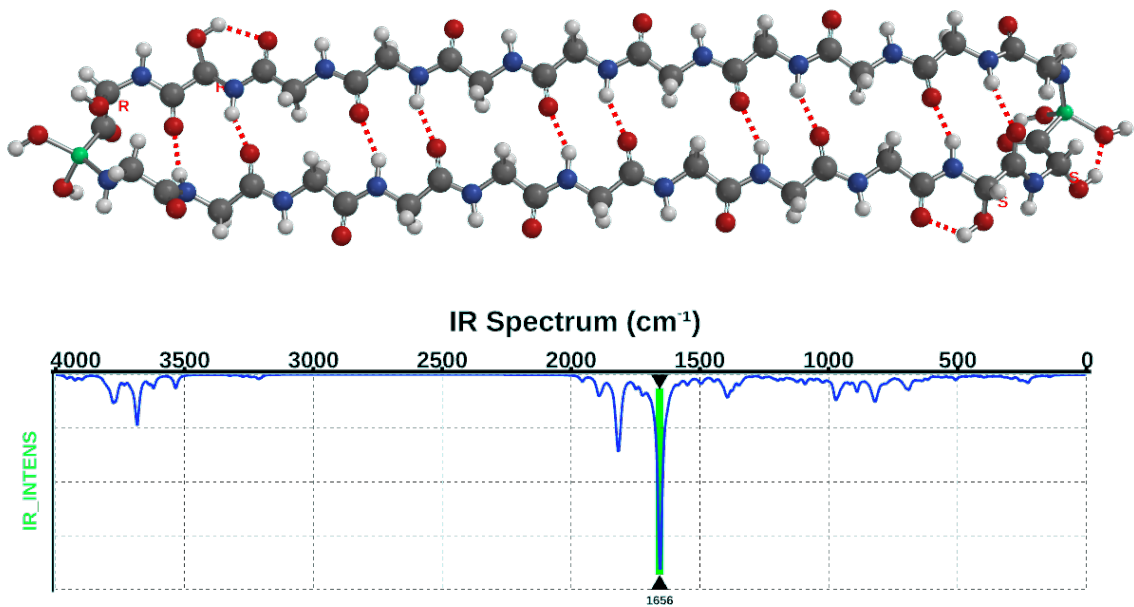

**Figure 4. IR absorption from the core of Glycine$_{18}$ Hydroxy-glycine$_4$ Fe$_2$O$_4$.** An absorption peak around 1651-8 cm$^{-1}$ (6.0 µm), shown here at 1656cm$^{-1}$ is the typical "amide I" absorption for the backbone of the anti-parallel beta sheet [8]. This IR spectrum is typical of the core and of the pared down conformers. Molecule format is ball and spoke. Atom labels: hydrogen white, carbon black, nitrogen blue, oxygen red, iron green.

**Discussion**
It is tempting to speculate on the consequences of a chirality-driven absorption in the visible region when the molecule in question shows evidence, through its great dominance, of some form of replication [4,5]. We consider the possible mechanisms that could benefit from such a specific source of energy in the field of a new star. One of these is energy transfer into vibrations.

**Electronic to vibrational energy transfer**
The electron promoted by 480nm absorption has an available energy of 2.58 eV that may be expended in chemical activity or vibrational excitation before it has a chance to radiatively decay in an estimated 300ns. If this electron moves down the molecular chain there is the possibility of energy transfer to collective vibrational modes. The electron wavelength $\lambda = h/p$ at 2.58eV is 0.764nm whereas the repeat length in the chain is $a = 0.718 \pm 0.004$nm, from the present calculations. When there is an exact match (resonance) between the electron wavelength and the period of the chain there is strong Bragg reflection of the electron and possible momentum transfer to the chain structure. A periodic potential along the chain, such as exists here, modifies the situation [14,15] by creating propagating electron states just below the resonance energy, which is 2.92eV for this chain, so our 2.58eV electron may be able to propagate while losing energy via scattering to collective vibrational modes. Without additional dissipation, as for instance in the gas phase, this vibrational energy will re-distribute slowly within the whole molecule, with continuing localized surging according to the Fermi simulation [16]. Sufficient energy will be available to "unzip" a series of hydrogen bonds spanning the length of the molecule, further discussed below.

**The possibility of replication**

The core hemoglycin molecule is present as the largest sub-component in all the major mass spectrometry peaks in an extensive analysis [4] of meteorite extracts, with support for its proposed structure in an X ray diffraction study [5]. It is found in different CV3 meteorites and its extreme dominance strongly suggests that it could have been formed in a process of molecular replication. Such a simple structure could easily replicate as a template on which further glycine chains could condense while being held in the correct orientation by hydrogen bonds at the outer edge of the parent molecule (Figure 5).

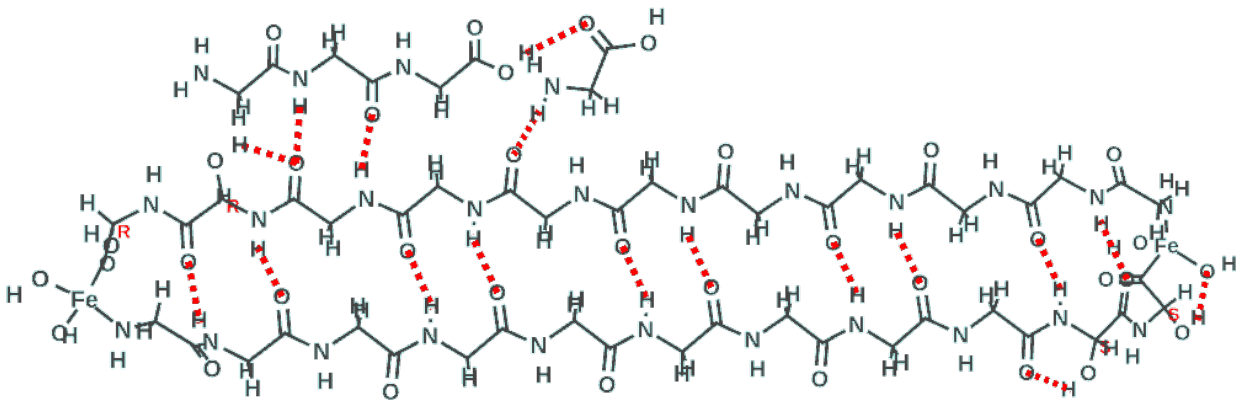

**FIGURE 5. Cartoon of "core" replication process. Line molecular format. A peptide of 4 glycine units has started a replication and aligned to the core with a single glycine coming in to form an amide bond to make a 5-mer glycine. Hydrogen bonds in red. R/S labeling denotes chirality of hydroxy-glycine units.**

The length of the newly-condensed poly-glycine chain would be determined by the original, and if Fe atoms were present its ends could be capped, followed by the rapid completion of a new anti-parallel chain, which would be exothermic [17]. The new molecule would then need to break free. A thermodynamic argument quantifies the energy required for replication [18]. In the present case the energy should at a minimum be sufficient to break that line of hydrogen bonds, which in calculation amounts to 71 kJ/mol [5] (0.74eV). The most likely energy source is visible radiation from a new star into a young circumstellar disc.

Of the three possible visible absorptions the strongest is the induced one at 480nm (Table 2) while lesser absorptions exist at 596nm and 401nm. The 596nm absorption is located at the Fe atom distal to the Fe atom with the 480nm absorption and is half as strong. The 401nm absorption is relatively very weak. In consequence, the induced 480nm absorption is a candidate for the energy source in the proposed hemoglycin replication. According to our finding, if vibrational energy transfer is strong, such replication could be chirally biased toward hemoglycin that contained R-$G_{OH}$ due to the 480nm absorption that is induced by R- $Gly_{OH}$ on the C-terminal. In the future it is hoped to have sufficient sample size to measure the chirality of the 1494 polymer, or its subunits.

## Materials and Methods

Modeling of this array of molecules was performed using Spartan '20 software [19], which incorporates a versatile and informative graphical user interface, executing calculations from the molecular modeling (MMFF) level up to advanced quantum calculations at the RPA level in embedded Q-chem [20]. Calculations were run on an M1 chip (Macbook Pro 17.1 with 8-core CPU used in 8 parallel threads). As the hemoglycin core unit had 780 active electrons in *ab initio* calculations at the 3-21G* level, it represented the practical upper limit for this computer and we were helped to run a longer glycine unit polymer calculation on $Glycine_{22}$ $Hydroxy-glycine_4$ $Fe_2O_4$ by Spartan Wavefunction chemists [19] who could employ greater computing capacity for convergence. Absorption wavelengths and strengths were calculated to 20 excited states, which covered the range down to the mid-ultraviolet. Transition strengths were typically plotted on a $log_{10}$ vertical scale unless mentioned otherwise, and an artificial width of 40nm was applied to the spectral peaks to simulate the typical molecular broadening via vibrations. The calculations were all in gas phase, which we believe is more appropriate for molecules in molecular clouds or discs.

# McGeoch
# S1 Supplementary Material

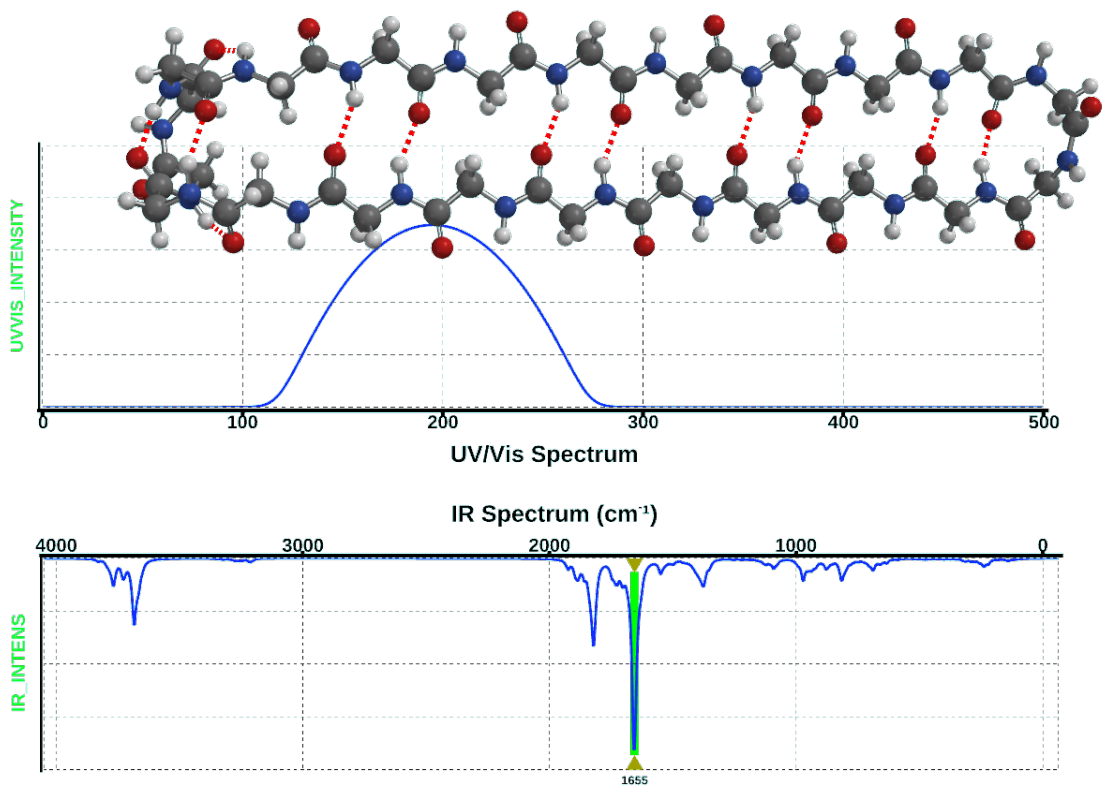

**S1 Figure 1.** UVvis and IR absorption from a loop of 22 glycine units in nm. A loop with no Fe atoms typically absorbs from 150-200nm, shown here at 191.43nm. The IR high-lit peak at 1658cm$^{-1}$ (6µm) is from the amide backbone. The molecular model format is ball and spoke. Atom labels: hydrogen white, carbon black, nitrogen blue, oxygen red.

**Movie to be added separately**

**S1 Figure 2.** Movie of the strong 3896cm$^{-1}$ (2.5μm) IR vibration of the rectus OH group on a glycine unit at the C-terminus of the core polymer: Glycine$_{18}$ Hydroxy-glycine$_4$ Fe$_2$O$_4.$ Molecule format is ball and spoke. Atom labels: hydrogen white, carbon black, nitrogen blue, oxygen red, iron green.

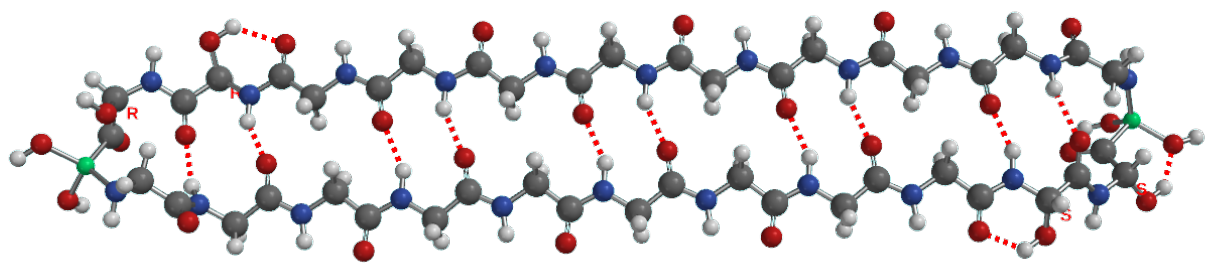

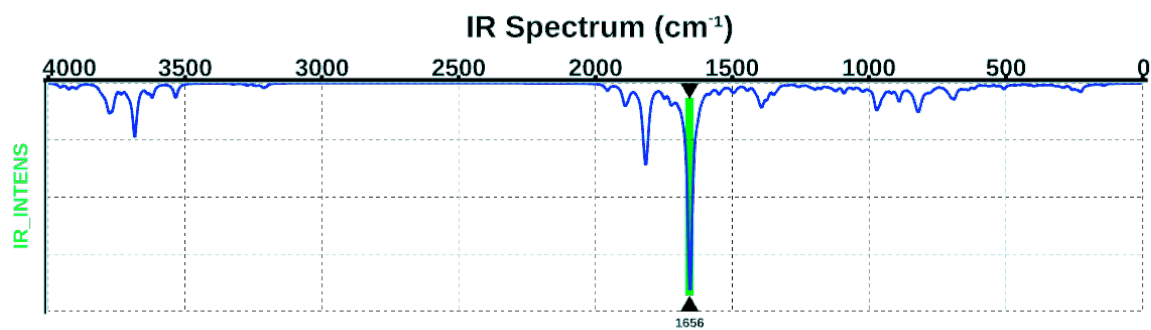

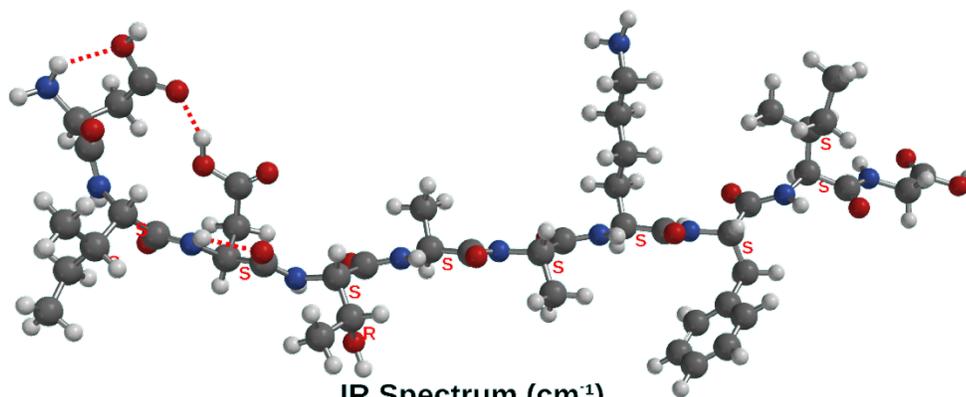

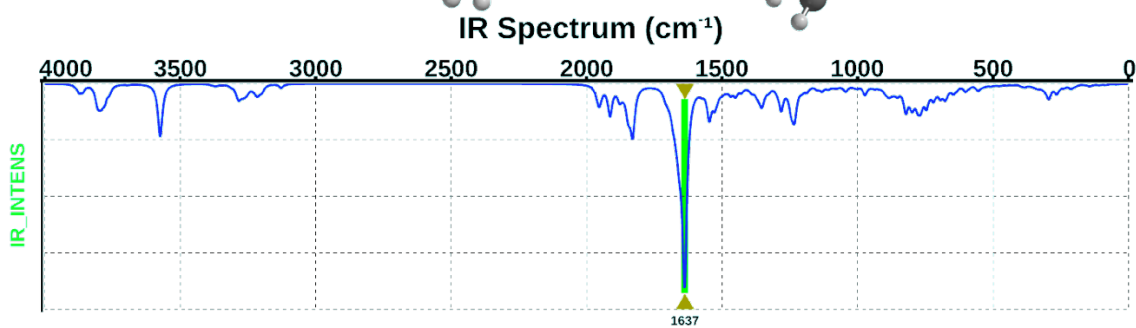

**S1. Figure 3.** A comparison of the IR absorption for the amide backbone of the "Core" space polymer with the 1st 10 amino acids of Subunit C of ATP synthase (DIDTAAKFIG). The subunit C molecule is depicted as a beta sheet for an IR absorption comparison to the Core space polymer. The amide backbone of both absorbs at 1635-1660cm-1 (region of 6μm). Molecule format is ball and spoke. Atom labels: hydrogen white, carbon black, nitrogen blue, oxygen red, iron gre

14